\newcount\style \style=1
\ifodd\style
\documentstyle[12pt,epsf,rotate,aaspp4]{article}
\fi 

\begin{document}

\newcommand\bb[1] {   \mbox{\boldmath{$#1$}}  }

\newcommand\del{\bb{\nabla}}
\newcommand\bcdot{\bb{\cdot}}
\newcommand\btimes{\bb{\times}}
\newcommand\vv{\bb{v}}
\newcommand\B{\bb{B}}
\newcommand\BV{Brunt-V\"ais\"al\"a\ }
\newcommand\iw{ i \omega }
\newcommand\kva{ \bb{k\cdot v_A}  }
\newcommand\kb{ \bb{k\cdot b}  }
\newcommand\kkz { \left( \frac{k}{k_Z}\right)^2\>}

    \def\dd{\partial}
    \def\tilde{\widetilde}
    \def\etal{et al.}
    \def\eg{e.g. }
    \def\etc{{\it etc.}}
    \def\ie{i.e.}
    \def\beq{ \begin{equation} }
    \def\eeq{ \end{equation} }
    \def\spose#1{\hbox to 0pt{#1\hss}} 
    \def\ltsim{\mathrel{\spose{\lower.5ex\hbox{$\mathchar"218$}}
         \raise.4ex\hbox{$\mathchar"13C$}}}

\def\tilde{\widetilde}

\newcommand{\schwz}{ {\dd  \ln P\rho ^{-\gamma} \over \dd Z}}
\newcommand{\schwR} { {\dd  \ln P\rho ^{-\gamma} \over \dd R} }

\newcommand{\balbz}{ {\dd  \ln T \over \dd Z}}
\newcommand{\balbR} { {\dd  \ln T \over \dd R} }

\long\def\Ignore#1{\relax}

\title{Local Axisymmetric Diffusive Stability of Weakly-Magnetized,
Differentially-Rotating, Stratified Fluids}

\author{Kristen Menou\altaffilmark{1,2} and Steven A.~Balbus}
\affil{Virginia Institute of Theoretical Astronomy, Department of Astronomy,}
\affil{University of Virginia, Charlottesville, VA 22903, USA.}

\and
\author{Henk C. Spruit}
\affil{Max Planck Institute for Astrophysics, Box 1317, 85741 Garching, Germany.}

\altaffiltext{1}{Celerity Foundation Fellow}

\altaffiltext{2}{Current Address: Institut d'Astrophysique de Paris,
98bis Boulevard Arago, 75014 Paris, France}

\begin{abstract}
We study the local stability of stratified, differentially-rotating
fluids to axisymmetric perturbations in the presence of a weak
magnetic field and of finite resistivity, viscosity and heat
conductivity. This is a generalization of the
Goldreich-Schubert-Fricke (GSF) double-diffusive analysis to the
magnetized and resistive, triple-diffusive case. Our fifth-order
dispersion relation admits a novel branch which describes a magnetized
version of multi-diffusive modes. We derive necessary conditions for
axisymmetric stability in the inviscid and perfect-conductor
(double-diffusive) limits. In each case, rotation must be constant on
cylinders and angular velocity must not decrease with distance from
the rotation axis for stability, irrespective of the relative strength
of viscous, resistive and heat diffusion. Therefore, in both
double-diffusive limits, solid body rotation marginally satisfies our
stability criteria. The role of weak magnetic fields is essential to
reach these conclusions. The triple-diffusive situation is more
complex, and its stability criteria are not easily stated.  Numerical
analysis of our general dispersion relation confirms our analytic
double-diffusive criteria, but also shows that an unstable
double-diffusive situation can be significantly stabilized by the
addition of a third, ostensibly weaker, diffusion process.  We
describe a numerical application to the Sun's upper radiative zone and
establish that it would be subject to unstable multi-diffusive modes
if moderate or strong radial gradients of angular velocity were
present.

\end{abstract}


\section{Introduction}

Developments in the last decade have made it clear that magnetic
fields, even weak magnetic fields, are essential to our understanding
of the dynamics of differentially rotating accretion disks and accretion
flows in general (Balbus \& Hawley 1991; 1998; Balbus 2003; Blaes 2003).
Although Keplerian rotation profiles are linearly stable to hydrodynamical
axisymmetric perturbations, the introduction of a weak magnetic field,
acting as a tether between fluid elements, renders such disks unstable to
the magnetorotational instability, or MRI.  The ensuing MHD turbulence
is now generally viewed as the primary mechanism providing the outward
angular momentum transport responsible for accretion in sufficiently
ionized, non self-gravitating disks (Hawley, Gammie \& Balbus 1995;
Armitage 1998; Hawley 2001; Balbus \& Hawley 1998).

The destabilizing effect on stellar differential rotation of an
embedded magnetic field was studied well before the accretion disk
community took notice of its importance.  Fricke (1968) pointed out,
for example, that time-steady field configurations (isovelocity
rotation contours running along field lines) need not be stable.
Acheson's thoroughgoing and detailed review (1973) analyzed the
effects of toroidal fields, noting that even very weak fields
qualitatively change the Goldreich-Schubert-Fricke (Goldreich \&
Schubert 1967; Fricke 1968) criterion for rotational stability, a
topic with which we shall be very much concerned in this paper.  The
magnetic field problem in its full generality was avoided, however,
because of the apparent complications associated with a time-dependent
equilibrium field caused by shear.  The clear understanding that very
general magnetic field configurations that are inconsequential for the
equilibrium state can have profound consequences for local WKB
perturbations is one of the key conceptual points that emerged in the
first accretion disk studies of the MRI (Balbus \& Hawley 1991, 1992).
In this weak field limit, the time dependence of the unperturbed
magnetic field is a nonissue, which opens a broad range of problems to
analysis.

The literature on the hydrodynamical stability of stellar differential
rotation is vast.  The magneto-hydrodynamical stability of differential
rotation is a much more specialized topic (e.g., Mestel 1999), and has
naturally tended to emphasize the direct dynamical forces associated
with the field itself.  The Sun and many other stars are expected to
possess magnetic fields buried deep inside their radiative zones, as
remnants of their complex formation history.  Their strength is not
well known.  It is well known, however, that the time for buried
magnetic fields to diffuse out of the solar interior is very long
indeed (see, e.g., Parker 1979).  Balbus \& Hawley (1994) and Balbus
(1995; hereafter B95) have studied the linear, adiabatic MRI in a
stably-stratified stellar system, and have noted that the
strongly-restoring buoyant forces limit the instability to
displacements lying only within spherical shells.  The \BV frequency
$N$ in stars is generally orders of magnitude larger than the rotation
frequency $\Omega$ (a value of $10^3$ is typical of the solar
radiative zone), whereas $N$ is at best comparable to $\Omega$ in a
disk.  Even more importantly, the most unstable adiabatic MRI modes in
a disk are always in the mid-plane (this maximizes the component of the
displacement along the angular velocity gradient), and these are
insensitive to the vertical stratification profile.

Goldreich \& Schubert (1967; hereafter GS) showed that the stabilizing
effects of entropy stratification can be compromised by thermal diffusion.
The mechanism is analogous to ``salt-fingering'' in the oceans.  In this
process, warm salty water overlying cool fresh water, naively a stable
configuration, is destabilized by heat transfer.  Warm fingers of salty
water, penetrating into the cooler waters below, diffuse heat outward more
rapidly than they diffuse salt, and thereby loose their buoyancy.  In the
stellar case, a downwardly displaced fluid element is adiabatically heated
and is normally warmer than its ambient surroundings.  This results in a
restoring buoyant force.  But if thermal diffusion causes sufficiently
rapid heat leakage, the buoyant force is diminished, and destabilizing
angular momentum gradients are then able to operate.  GS found that
not only must the familiar Rayleigh stability criterion of increasing
angular momentum with increasing axial radius be satisfied, the presence
of a large thermal conductivity implies that the angular velocity must
also be constant along cylindrical axes---a far more stringent criterion.

But great care must therefore be given to apparently small
diffusivities when assessing the stability of a stellar rotation
profile.  The GS result holds when the ratio of the thermal to viscous
diffusivities is sufficiently large.  Large compared to what?  The
answer is not unity (Acheson 1978).  Rather, it must be large compared
to the square of the ratio of the \BV to rotation frequency, a
condition not met for the solar radiative zone. Acheson (1978) also
generalized the study of multi-diffusive modes to non-axisymmetric
perturbations in a medium with a purely toroidal magnetic field, and
found that the angular velocity gradient, not the angular momentum
gradient, emerged as the rotational stability discriminant.  This
result, it turns out, is very general, extending beyond toroidal field
geometries (Balbus 1995, 2001).

In this paper, we solve the problem of multi-diffusive stability to
axisymmetric perturbations in the presence of rotation, entropy
stratification, and a magnetic field of arbitrary geometry.  We find
that the results are sensitive to the triple combination of
resistivity, viscosity, and thermal conductivity, and that all three
must be included in the analysis from the very beginning. In what
follows, we will refer to such a situation as triple-diffusive,
whether it is stable or not.

An outline of the paper is as follows.  Section 2 is a derivation of the
general dispersion relation described above.  Section 3 is a detailed
stability analysis.  Section 4 applies the results specifically to the
sun's radiative interior.  A more general discussion follows in \S 5,
and \S 6 summarizes our conclusions.  

\section{Dispersion Relation}

The MHD equations including the effects of viscosity, resistivity and
heat conduction take the form

\beq
{\dd\rho\over \dd t} + \del\bcdot (\rho \bb {v}) =  0,
\eeq
\beq\label{mom}
\rho {\dd\vv \over \dd t} + (\rho \vv\bcdot\del)\vv = -\del\left(
P + {B^2\over 8 \pi} \right)-\rho \del \Phi +
\left( {\B\over 4\pi}\bcdot \del\right)\B + \mu \left( \del^2 \vv + 
\frac{1}{3} \del(\del \bcdot \vv) \right),
\eeq
\beq
{\dd\bb{B}\over \dd t} = \bb {\nabla\times (v\times B)} - 
  \eta  { \bb \nabla\times} \left( {\bb \nabla\times \bb{B}} \right),
\eeq
\beq
{1\over (\gamma -1)} {P} {d\ln P\rho^{-\gamma}\over dt} = \chi \del^2 T.
\eeq

\noindent These are, respectively, the continuity equation, the
momentum conservation equation, the induction equation and the
entropy-form of the energy equation (see, e.g., Balbus \& Hawley
1998). Our notation is as follows: $\bb{v}$ is the flow velocity,
$\rho$ is the mass density, $P$ is the pressure, $\bb{B}$ is the
magnetic field, $\Phi$ is the gravitational potential, $T$ is the
temperature, $\mu$ is the dynamic viscosity, $\eta$ is the resistivity
and $\chi$ is the heat conductivity (which can represent thermal or
radiative conductivity, depending on the problem at hand). In what
follows, we write the kinematic viscosity coefficient
$\nu=\mu/\rho$. Bulk viscosity effects are neglected. The adiabatic
index of the gas, denoted $\gamma$, is $5/3$ for a monotomic gas with
negligible radiation pressure.  {We have ignored the spatial
dependence of the diffusion coefficients $\mu$, $\eta$ and $\chi$,
which is appropriate for a leading order WKB analysis.  We have also
neglected the resistive and viscous dissipation terms in the entropy
equation, which are also higher order terms. The validity of this
approximation is examined in Appendix~A.}

We work in cylindrical coordinates $({R}, {\phi}, {Z})$. We consider
axisymmetric Eulerian perturbations (denoted by a prefix $\delta$)
with WKB space-time dependence $\exp [i(\bb{k\cdot r} - \omega t)]$,
where $\bb{k}= (k_R, 0, k_Z)$. The basic state magnetic field, allowed
to have any geometry, is assumed to be weak enough that it does not
affect the basic state configuration, i.e. weak compared to both
rotation and pressure gradients.  The basic state rotation is given by
$\bb{\Omega} = (0, 0, \Omega(R,Z))$ along the ${Z}$--axis. We neglect
in our analysis weak circulations such as those induced by a finite
viscosity or any meridional circulation (when considering stellar
applications).

Using the Boussinesq approximation, the leading order WKB terms become
after linearization
\beq\label{divv}
k_R \delta v_R + k_Z \delta v_Z =0,
\eeq
\begin{eqnarray}
\left(-\iw + \nu k^2\right) \delta v_R && +{i k_R \over \rho} \delta P - 2 \Omega \delta
v_\phi
- {\delta\rho\over\rho^2} { \dd P \over \dd R} + {i k_R \over 4
\pi
\rho}\nonumber\\
&&\times \left( B_\phi \,\delta B_\phi + B_Z \,\delta B_Z
\right) - {ik_Z\over
4 \pi\rho}B_Z\, \delta B_R = 0,
\end{eqnarray}
\beq
\left(-\iw  + \nu k^2\right) \delta v_\phi + \delta v_R\, {1\over R}{\dd (R^2\Omega) \over \dd R}+
\delta v_Z\, R{\dd \Omega\over \dd Z}
-i \bb{k \cdot B} {\delta B_\phi \over 4 \pi \rho} = 0,
\eeq
\begin{eqnarray}
\left(-\iw  + \nu k^2\right) \delta v_Z &&+ {i k_Z \, \delta P \over \rho} -
{\delta\rho\over
\rho^2}{ \dd P \over \dd Z} + {i k_Z \over 4 \pi \rho
}\nonumber\\
&& \times \left( B_\phi \, \delta B_\phi
+ B_R\, \delta B_R \right) - {ik_R B_R\over 4 \pi \rho}\, \delta
B_Z =0,
\end{eqnarray}
\beq\label{indR}
\left(-\iw + \eta k^2\right) \delta B_R - i \bb{k \cdot B} \delta v_R = 0,
\eeq
\beq
\left(-\iw + \eta k^2\right) \delta B_\phi - \delta B_R\, {\dd\Omega\over \dd \ln R} -
\delta
B_Z\, R {\dd\Omega\over \dd Z} - i \bb{k\cdot B} \, \delta
v_\phi = 0,
\eeq
\beq\label{indZ}
\left(-\iw  + \eta k^2\right) \delta B_Z - i \bb{k\cdot B} \delta v_Z = 0,
\eeq
\beq\label{lincon}
\iw {\gamma} {\delta\rho\over \rho} +
\bb{(\delta v\cdot\nabla)}\ln P\rho^{-\gamma} =
(\gamma-1){\chi T k^2 \over P}  {\delta \rho\over\rho} .
\eeq
\noindent The gravitational potential has dropped from the problem
because we are interested in wavelengths much shorter than the Jeans
wavelength at which self-gravity would become important.  The relation
$\delta T= -T(\delta \rho / \rho)$ has been used in
equation~(\ref{lincon}) above.

Solving for the eight $\delta$--unknowns in these eight equations, we
obtain the following dispersion relation (in compact form)
\begin{eqnarray}
{\tilde\omega_{b+v}}^4 {\omega_{e}}\, {k^2\over k_Z^2} + {\tilde\omega_{b+v}}^2 {\omega_{b}}
\left[ {1\over \gamma \rho}\, \left({\cal D} P\right)\, {\cal D} \ln
P\rho^{-\gamma}\right]
+ {\tilde\omega_{b}}^2 {\omega_{e}} \left[
{1\over R^3}\, {\cal D} (R^4\Omega^2) \right] - 4 \Omega^2 (\kva)^2 {\omega_{e}}=
0,
\end{eqnarray}
where
\begin{eqnarray}
\bb{v_A} = { \bb{B}/\sqrt{4\pi\rho}}, \qquad 
k^2 = k_R^2 + k_Z^2, \qquad {\tilde\omega_{b+v}}^2=\omega_b \omega_v -(\kva)^2,
\qquad {\tilde\omega_{b}}^2=\omega_b^2 -(\kva)^2, \nonumber
\end{eqnarray}
\begin{eqnarray}
\omega_b=\omega+i \eta k^2, \qquad  \omega_v=\omega+i \nu k^2, \qquad
\omega_e=\omega+ \frac{\gamma -1}{\gamma}{i  T\over  P} \chi k^2, 
\qquad {\cal D} \equiv \left( \frac{k_R}{k_Z}\frac{\dd}{\dd Z}
-\frac{\dd}{\dd R}\right) \nonumber
\end{eqnarray}

\noindent By substituting $\sigma =-i \omega$, this 5th order dispersion
relation can be written in developed form

\beq\label{poly}
a_0\sigma^5 + a_1\sigma^4 + a_2\sigma^3 + a_3\sigma^2 + a_4\sigma +a_5 =0,
\eeq
where
\beq
a_0 = k^2/k_Z^2,
\eeq
\beq
a_1 = \frac{k^2}{k_Z^2} \left[ 2\nu k^2  + 2\eta k^2  + \xi k^2  \right],
\eeq
\begin{eqnarray}
a_2 = \frac{k^2}{k_Z^2} \left[ \nu^2 k^4 + \eta^2 k^4 +4\nu\eta k^4  +2\nu\xi k^4 
 +2\eta\xi k^4  + 2(\bb{k\cdot v_A})^2   \right] \nonumber\\
-\left[ \frac{1}{\gamma \rho}{\cal D}P\>  {\cal D}\ln P\rho^{-\gamma} \right]
- \left[ \frac{1}{R^3}{\cal D} (R^4\Omega^2) \right],
\end{eqnarray}
\begin{eqnarray}
a_3 = \frac{k^2}{k_Z^2} \left[ 2\eta\nu^2 k^6 + 2\nu\eta^2 k^6  +\nu^2\xi k^6 
+\eta^2\xi k^6+4\nu\eta\xi k^6 +2(\nu k^2+\eta k^2+\xi k^2)(\bb{k\cdot v_A})^2\right]
\nonumber\\
-(2\eta k^2+\nu k^2)\left[\frac{1}{\gamma \rho}{\cal D}P\>  {\cal D}\ln P\rho^{-\gamma}\right]-
(2\eta k^2+\xi k^2)\left[\frac{1}{R^3}{\cal D} (R^4\Omega^2)\right],
\end{eqnarray}
\begin{eqnarray}
a_4 = \frac{k^2}{k_Z^2} \left[ 2\eta\xi\nu^2 k^8 + 2\nu\eta^2\xi k^8  +\eta^2\nu^2 k^8
+2(\nu\eta k^4+\nu\xi k^4 +\eta\xi k^4)(\bb{k\cdot v_A})^2 +(\bb{k\cdot v_A})^4 \right]
\nonumber\\
-(2\nu\eta k^4+\eta^2 k^4+(\bb{k\cdot v_A})^2)\left[\frac{1}{\gamma \rho}{\cal D}P\>  
{\cal D}\ln P\rho^{-\gamma}\right] \nonumber\\
-(2\eta\xi k^4+\eta^2 k^4+(\bb{k\cdot v_A})^2)\left[\frac{1}{R^3}{\cal D} 
(R^4\Omega^2)\right] 
\nonumber\\
-4\Omega^2(\bb{k\cdot v_A})^2,
\end{eqnarray}
\begin{eqnarray}
a_5 = \frac{k^2}{k_Z^2} \left[ \xi\eta^2\nu^2 k^{10} + 
2\xi\nu\eta k^{6}(\bb{k\cdot v_A})^2 + \xi k^{2}(\bb{k\cdot v_A})^4 \right]
\nonumber\\
-(\nu\eta^2 k^{6}+\eta k^{2}(\bb{k\cdot v_A})^2)\left[\frac{1}{ \gamma \rho}{\cal D}P\>  
{\cal D}\ln P\rho^{-\gamma}\right]
\nonumber\\ 
-(\xi\eta^2 k^{6}+\xi k^{2}(\bb{k\cdot v_A})^2)\left[\frac{1}{R^3}{\cal D} 
(R^4\Omega^2)\right] \nonumber\\
-4\Omega^2(\bb{k\cdot v_A})^2\xi k^{2},
\end{eqnarray}

\noindent where, for conciseness, we have introduced the heat
diffusivity
\begin{eqnarray}
\xi=\frac{\gamma-1}{\gamma} \frac{T}{P} \chi. \nonumber
\end{eqnarray}

The above dispersion relation reduces to several previously
established relations in the appropriate limits.  Taking
$\nu=\eta=\chi=0$ (diffusion-free limit), one recovers the result of
B95 (his Eq.~[2.4]), which is a generalization of the work of Balbus
\& Hawley (1991; 1994) on the MRI to general rotation laws
(i.e. non-constant on cylinders). Taking $v_A=\eta=0$ (hydrodynamic
limit), one recovers exactly the result of Goldreich \& Schubert
(1967; their Eq.~[32]; see also Fricke 1968). Finally, taking
$\nu=\eta=0$ but $\chi$ finite, one recovers the result of Balbus
(2001; eq.~[29]) provided that the isotropic thermal conductivity
limit is substituted in his dispersion relation.  This is effected by
setting $(\bb{k\cdot b})^2 \rightarrow k^2$ and ${\cal D} \ln T
\rightarrow 0$ in the Balbus (2001) dispersion relation.  (See also
Urpin \& Brandenburg 1998).

\section{Stability Analysis}

The complexity of our dispersion relation makes it difficult to derive
general necessary {\it and} sufficient conditions for stability.  The
single-diffusive case (Balbus 2001) is amenable to a Routh-Hurwitz
(RH) analysis, but we have found this method impractical for the
multi-diffusive case of interest here. Instead, we derive a series of
necessary conditions for stability. As we shall see, they are
stringent enough to provide very useful limits on the maximum
allowable level of stable differential rotation.

To simplify the analysis further, we first consider the inviscid and
perfect-conductor limits separately. This allows us to reduce the
problem to two separate double-diffusive situations. As we shall see
below, this separation has to be made with some caution.  Our
stability analysis then proceeds as follows. Given that $a_0 > 0$ in
Eq.~[\ref{poly}], if any of the five other $a_i$ coefficients is
negative, there will be at least one unstable root (i.e. one with a
strictly positive real part) to the dispersion relation. A necessary
condition for stability is thus that all the $a_i$ be positive. The
requirement $a_1 > 0$ is trivially satisfied, so we focus on the
criteria corresponding to the positivity of the last four coefficients
below. Notice that all the coefficients $a_2$--$a_5$ possess first
terms in bracket, $\propto k^2/k_Z^2$, that are strictly
positive. These terms represent the systematically stabilizing effects
of diffusion processes, ($\nu$, $\eta$, $\chi$), and magnetic tension,
$(\bb{k\cdot v_A})^2$, on very small scales, where they become
dominant. Quite generally, we can focus our attention on (at least
somewhat) larger scales for which the coefficients $a_2$--$a_5$ depend
on terms combining the effects of differential rotation,
stratification, diffusion and magnetic tension, and for which
positivity is not guaranteed.

In addition, we have found that it is possible to carry out a partial
RH analysis. The first non-trivial determinant required to be positive
by the RH criterion is $det(2)=a_1a_2 - a_0a_3 > 0$.  This comes as an
additional necessary condition for stability. We have found that this
requirement is important to our conclusions on differential rotation.

\subsection{Perfect-Conductor Limit ($\eta \rightarrow 0$)}

This is the case closest to the analysis of Goldreich \& Schubert
(1967) and Fricke (1968), who focused on double-diffusive
hydrodynamical stability in the presence of viscosity and heat
diffusion.

\subsubsection{Requirement $a_2 > 0$ for stability when $\eta \rightarrow 0$}

>From the structure of the coefficient $a_2$ (involving pure rotational
and stratification terms), we recognize a stability criterion related
to diffusion-free, hydrodynamical modes.  On large enough scales 
for the stabilizing effect of the first bracket term in
$a_2$ to be unimportant, the criterion $a_2 > 0$ becomes
\beq -\left[ \frac{1}{\gamma \rho}{\cal D}P\> {\cal D}\ln
P\rho^{-\gamma} \right] - \left[ \frac{1}{R^3}{\cal D} (R^4\Omega^2)
\right] >0.  \eeq
As discussed by B95, this
inequality translates into the classical Solberg-H\o iland criteria
(see, e.g., Tassoul 1978)

\beq\label{hoilad1}
-{1 \over \gamma\rho}(\del P)\bcdot\del\ln P\rho^{-\gamma}
+ {1\over R^3} {\dd R^4\Omega^2\over \dd R} > 0,
\eeq
\beq\label{hoilad2}
\left( - {\dd P\over \dd Z} \right) \, \left( {1 \over R^3}
{\dd R^4 \Omega^2\over\dd R} \schwz - {1 \over R^3} 
{\dd R^4\Omega^2\over\dd Z}\schwR \right) > 0.
\eeq

\noindent For a star with spherically-symmetric isocontours of density, $\rho$,
and pressure, $P$, these criteria take a more familiar form when both
cylindrical ($R$, $Z$) and spherical (${r}$, ${\theta}$) coordinates
are used
\beq
N^2 + \kappa^2 > 0,
\eeq
\beq \label{GSshell}
N^2 \cot \theta \frac{\dd \left( \Omega \sin^2 \theta \right) }
{\dd \theta} 
> 0,
\eeq
where the square of the \BV frequency is defined by
\beq
N^2 =  -{1 \over \gamma\rho} {\dd  P \over \dd r} 
{\dd  \ln P\rho ^{-\gamma} \over \dd r}, 
\eeq
and the square of the epicyclic frequency is defined by 
\beq
\kappa^2 = {1\over R^3} {\dd R^4\Omega^2\over \dd R}.
\eeq

\noindent Equation~(\ref{GSshell}) shows that to avoid rotational
instability in a stratified star, the specific angular momentum must
not decrease away from the rotation axis within a spherical
shell. While these criteria are necessary and sufficient to guarantee
stability in the diffusion--free, hydrodynamical limit, they become
only necessary conditions for stability in the broader context studied
here.

\subsubsection{Requirement $a_3 > 0$ for stability when $\eta \rightarrow 0$}

>From the structure of the coefficient $a_3$, we recognize a stability
criterion related to hydrodynamical modes influenced on all scales by
viscosity and heat diffusion. This essentially is the constant term of
Goldreich \& Schubert's third--order dispersion relation.  On large
enough scales for the stabilizing effect of the first
bracket term to be unimportant, the criterion $a_3 > 0$ becomes
\beq \label{a3req}
-\epsilon_{\nu}  \left[ \frac{1}{\gamma \rho}{\cal D}P\>  
{\cal D}\ln P\rho^{-\gamma} \right] - 
\left[ \frac{1}{R^3}{\cal D} (R^4\Omega^2) \right] >0, 
\eeq
where the Prandtl number is defined by
\beq
\epsilon_{\nu} = \frac{\gamma \nu}{\gamma -1} \frac{P}{T \chi}.
\eeq
This condition can be rewritten
\begin{eqnarray}
\left( \frac{k_R}{k_Z} \right)^2 \epsilon_{\nu} N_Z^2+ \frac{k_R}{k_Z} \left[
{ \epsilon_{\nu} \over \gamma \rho} \left( {\dd P\over \dd Z}\schwR +  {\dd P\over \dd R}
 \schwz \right) -{1\over R^3} {\dd (R^4 \Omega^2) \over\dd Z} \right] 
\nonumber\\
+ \epsilon_{\nu} N_R^2 + {1 \over R^3} {\dd (R^4 \Omega^2) \over\dd R} >0,  
\end{eqnarray}
where
\beq
N_R^2= -{1 \over \gamma\rho} {\dd  P \over \dd R} 
{\dd  \ln P\rho ^{-\gamma} \over \dd R},
\qquad
N_Z^2= -{1 \over \gamma\rho} {\dd  P \over \dd Z} 
{\dd  \ln P\rho ^{-\gamma} \over \dd Z}. 
\eeq

\noindent As a quadratic in $x=k_R/k_Z$, this inequality is satisfied
if and only if (i) the sum of the constant coefficient and the
coefficient for $x^2$ is positive, and (ii) the discriminant of the
polynomial in $x$ is negative.  By using the vorticity relation for
the basic state (e.g. Tassoul 1978)
\beq
R \frac{\dd \Omega^2 }{\dd Z} = \frac{1}{\rho^2} \left(
\frac{\dd \rho}{\dd R}\frac{\dd P}{\dd Z} - 
\frac{\dd \rho}{\dd Z}\frac{\dd P}{\dd R} \right),
\eeq
to simplify condition (ii), the two criteria can be written
\beq
\epsilon_{\nu} (N_R^2 + N_Z^2) + \frac{1}{R^3}\frac{\dd (R^4 \Omega^2)}{\dd R} > 0, 
\eeq
\begin{eqnarray}
(1-\epsilon_{\nu})^2 \left( {1\over R^3} {\dd (R^4 \Omega^2) \over\dd Z} \right)^2
+{ 4 \epsilon_{\nu} \over \gamma \rho} {\dd P\over \dd Z} \left[ {1\over R^3} 
{\dd (R^4 \Omega^2) \over \dd R} \schwz 
- {1 \over R^3} {\dd (R^4 \Omega^2) \over\dd Z} \schwR \right] < 0.
\end{eqnarray}

\noindent For a star with spherically-symmetric isocontours of density, $\rho$,
and pressure, $P$, these criteria take the following form when both
cylindrical ($R$, $Z$) and spherical (${r}$, ${\theta}$) coordinates are used
\beq
\epsilon_{\nu} N^2 + \kappa^2 > 0,
\eeq
\beq 
(1-\epsilon_{\nu})^2 \left( R {\dd \Omega^2 \over\dd Z} \right)^2
-8 \epsilon_{\nu} N^2 \Omega \cot \theta \left[ {\dd (\Omega \sin^2 \theta) 
\over\dd \theta} \right] < 0.
\eeq

\noindent Note that in the limit $\epsilon_{\nu} \rightarrow 0$, these
two conditions reduce to the result found by Goldreich \& Schubert
(1967): necessary conditions for stability are (i) that the specific
angular momentum does not decrease with distance from the rotation
axis and (ii) that the rotation law be constant on cylinders. Acheson
(1978) has pointed out, however, that the limit $\epsilon_{\nu}
\rightarrow 0$ must be carefully taken because in some stars, in
particular the Sun, the product $\epsilon_{\nu} N^2$ is not
necessarily small compared to the rotational terms. We discuss this
limitation further in \S4. Interestingly, for $\epsilon_{\nu}=1$, the
diffusion-free result (\S3.1.1) is recovered.

\subsubsection{Requirement $a_4 > 0$ for stability when $\eta \rightarrow 0$}

>From the structure of the coefficient $a_4$ (note in particular the $-
4 \Omega^2 (\bb{k\cdot v_A})^2 $ term), we recognize a stability
criterion related to diffusion-free, MHD modes.  On large enough
scales  for the stabilizing effect of the first bracket
term to be unimportant, the criterion $a_4 > 0$ becomes

\beq
-\left[ \frac{1}{\gamma \rho}{\cal D}P\>  {\cal D}\ln P\rho^{-\gamma} \right]
- \left[ \frac{1}{R^3}{\cal D} (R^4\Omega^2) + 4 \Omega^2 \right] >0.
\eeq

\noindent Note the additional assumption made to derive this
inequality: the weak magnetic field must be strong enough that
magnetic tension forces are important on scales larger than those on
which dissipation stabilizes all perturbations (e.g. $(\kva)^2 \gg
\eta^2 k^4$). This still leaves a comfortable range of dynamically
interesting magnetic field strengths, as is shown for the specific
Solar case in Appendix~A.  As described by B95, the above inequality
translates into

\beq\label{hoilb1}
-{1 \over \gamma \rho}(\del P)\bcdot\del\ln P\rho^{-\gamma}
+ {\dd \Omega^2\over \dd \ln R} > 0,
\eeq
\beq\label{hoilb2}
\left( - {\dd P\over \dd Z} \right) \, \left(
{\dd \Omega^2\over\dd R} \schwz - {\dd \Omega^2\over\dd Z}\schwR
\right) > 0, 
\eeq

i.e. forms similar to the classical Solberg-H\o iland criteria but
with gradients of angular velocity replacing the traditional gradients
of specific angular momentum. For a star with spherically-symmetric
isocontours of density, $\rho$, and pressure, $P$, these criteria take
the following form when both cylindrical ($R$, $Z$) and spherical
(${r}$, ${\theta}$) coordinates are used

\beq
N^2 + {\dd \Omega^2\over \dd \ln R}  > 0,
\eeq
\beq
N^2 \sin \theta \cos \theta \frac{\dd \Omega}{\dd \theta} > 0.
\eeq

\noindent Note the important new element introduced by the presence of a
magnetic field: to guarantee axisymmetric stability in a
stably-stratified star, the angular velocity must not decrease with
distance from the rotation axis within a spherical shell. While the
two criteria above were necessary and sufficient conditions for
stability in the diffusion--free study of B95, they become
only necessary conditions for stability in the broader context studied
here.

\subsubsection{Requirement $a_5 > 0$ for stability when $\eta \rightarrow 0$}

>From the structure of the coefficient $a_5$, we recognize a stability
criterion related to MHD modes influenced by viscosity and heat
diffusion, not just on small scales.  On large enough scales for the
stabilizing effect of the first bracket term to be unimportant, the
criterion $a_5 > 0$ becomes

\beq
-\left[ \frac{1}{R^3}{\cal D} (R^4\Omega^2) + 4 \Omega^2 \right] >0. 
\eeq

\noindent Note the remarkable property of this stability condition: it
is independent of the stratification term (which drops out of the
dispersion relation in the limit $\eta \rightarrow 0$). (This property
would not be exactly conserved in a triple-diffusive system.)  To be
satisfied for any combination of $k_R$ and $k_Z$, it requires the
following two conditions

\beq
 \frac{\dd \Omega^2}{\dd \ln R} > 0, 
\eeq
\beq
\left( R {\dd \Omega^2 \over\dd Z} \right)^2 < 0,
\eeq

\noindent so that marginal stability is possible only for a rotation
law that is constant on cylinders ($\dd \Omega / \dd z = 0$).

In a realistic situation, however, even a fully ionized plasma will
possess a finite resistivity. One should include resistivity in the
analysis, no matter how small it is, because the stratification term
can generally be much larger than the rotational term in a stellar
context. The situation is intrinsically triple-diffusive because both
viscosity and resistivity affect the momentum of displaced fluid
elements.  This makes the analysis of this branch of the dispersion
relation more complicated, and it appears that a simple stability
criterion independent of $k$ and $v_A$ does not exist in general.

\subsubsection{Requirement $det(2) > 0$ for stability when $\eta \rightarrow 0$}

Necessary conditions for stability can be made more stringent by the
additional requirement $det(2)= a_1a_2 - a_0a_3 > 0$. This is one of
the five RH determinants that must be positive.

Keeping all the terms in the coefficients $a_0$ to $a_3$ when
calculating $det(2)$, we note that several are strictly positive and
become negligibly small on large enough scales. The condition $det(2)
> 0$ can thus be reduced to

\beq \label{det2}
-(1+\epsilon_{\nu})  \left[ \frac{1}{\gamma \rho}{\cal D}P\>  
{\cal D}\ln P\rho^{-\gamma} \right] - 2 \epsilon_{\nu} 
\left[ \frac{1}{R^3}{\cal D} (R^4\Omega^2) \right] >0. 
\eeq

\noindent Note the different structure of this inequality as compared
to Eq.~(\ref{a3req}). The factor $\epsilon_{\nu}$ is now in front of
the rotational term (second bracket).  In general, for stars,
rotational effects are weak compared to the entropy stratification
term (first bracket) and $\epsilon_{\nu} \ll 1$, so that one is
tempted to reduce the above expression to the stratification term
only. This, however, is incorrect because terms of order $N^2 /
\Omega^2$ emerge, and are important. Using the equivalence of
Eq.~(\ref{det2}) with Eq.~(\ref{a3req}), if $\epsilon_{\nu}$ is
replaced by $(1+\epsilon_{\nu})/2\epsilon_{\nu}$ in the latter, the
analysis proceeds as in \S3.1.2 and we obtain the following two
necessary conditions for stability

\beq
(1+\epsilon_{\nu}) (N_R^2 + N_Z^2) + 
\frac{2 \epsilon_{\nu}}{R^3}\frac{\dd (R^4 \Omega^2)}{\dd R} > 0, 
\eeq

\begin{eqnarray}
(1-\epsilon_{\nu})^2 \left( {1\over R^3} {\dd (R^4 \Omega^2) \over\dd Z} \right)^2
+{ 8 \epsilon_{\nu} (1+\epsilon_{\nu}) \over \gamma \rho} {\dd P\over \dd Z} \left[ {1\over R^3} 
{\dd (R^4 \Omega^2) \over \dd R} \schwz 
- {1 \over R^3} {\dd (R^4 \Omega^2) \over\dd Z} \schwR \right] < 0.
\end{eqnarray}

\noindent For a star with spherically-symmetric isocontours of density, $\rho$,
and pressure, $P$, these criteria take the following form when both
cylindrical ($R$, $Z$) and spherical (${r}$, ${\theta}$) coordinates
are used

\beq
(1+\epsilon_{\nu}) N^2 + 2 \epsilon_{\nu} \kappa^2 > 0,
\eeq
\beq 
(1-\epsilon_{\nu})^2 \left( R {\dd \Omega^2 \over\dd Z} \right)^2
-16 \epsilon_{\nu} (1+\epsilon_{\nu}) N^2 \Omega \cot \theta 
\left[ {\dd (\Omega \sin^2 \theta) \over\dd \theta} \right] < 0.
\eeq

\noindent Note that these conditions differ from those obtained by requiring
that $a_3 > 0$ in \S~3.1.2. The first of the two conditions above
requires, in the limit $\epsilon_\nu \rightarrow 0$, that the
stratification be stable, this time independently of the rotational
stability (measured by $\kappa^2$). The second of the two conditions
above is more stringent than the corresponding one in \S~3.1.2 because
it involves a first term that cannot be negative (or, equivalently, is
stabilizing). Only the second term can, and its absolute value is a
factor of two larger in the above inequalities as compared to those we
derived in \S3.1.2.

It is worth emphasizing that the combination of requirements $a_2 > 0$
(\S~3.1.1), $a_3 > 0$ (\S 3.1.2) and $det(2)>0$ (above) constitute
necessary {\it and} sufficient stability conditions for the
double-diffusive hydrodynamical problem considered by Goldreich \&
Schubert (1967) and Fricke (1968). One easily shows that an
hydrodynamical system satisfying all these conditions also satisfies
the RH criterion for the third order dispersion relation of the purely
hydrodynamical problem.

\subsection{Inviscid Limit ($\nu \rightarrow 0$)}

For the most part, the stability analysis in the limit $\nu
\rightarrow 0$ proceeds in a manner very similar to the limit $\eta
\rightarrow 0$. Unless complications arise, we directly list the
necessary conditions for stability in a convenient form.

\subsubsection{Requirement $a_2 > 0$ for stability when $\nu \rightarrow 0$}

Like in the perfect-conductor limit, the necessary conditions for
stability are equivalent to the classical Solberg-H\o iland criteria
(see \S3.1.1).

\subsubsection{Requirement $a_3 > 0$ for stability when $\nu \rightarrow 0$}

On large enough scales, the requirement $a_3 > 0$ becomes, in this
case,

\beq
-2 \epsilon_{\eta} \left[ \frac{1}{\gamma \rho}{\cal D}P\> {\cal D}\ln
P\rho^{-\gamma} \right] - \left( 2 \epsilon_{\eta} +1 \right) \left[
\frac{1}{R^3}{\cal D} (R^4\Omega^2) \right] >0,
\eeq

\noindent where the ``Acheson number\footnote{We were not able to find
an existing terminology for this dimensionless number in the
literature. Acheson (1978) appears to have been the first to recognize
the importance of this dimensionless quantity for the problem of
differential rotation in magnetized and stratified fluids. This number
can be expressed, in a rather indirect way, as the ratio of the
Prandtl number to the magnetic Prandtl number.}''  $\epsilon_{\eta}$
is

\beq \label{Ache}
\epsilon_{\eta}=\frac{\gamma \eta}{\gamma -1}\frac{P}{T \chi}.
\eeq

\noindent By similarity with the analysis in \S3.1.2, we deduce the
following necessary conditions for stability

\beq
\frac{2 \epsilon_{\eta}}{1+2 \epsilon_{\eta}} 
(N_R^2 + N_Z^2) + \frac{1}{R^3}\frac{\dd (R^4 \Omega^2)}{\dd R} > 0, 
\eeq

\begin{eqnarray}
&& (1 - \frac{2 \epsilon_{\eta}}{1+2 \epsilon_{\eta}})^2 \left( {1\over R^3} 
{\dd (R^4 \Omega^2) \over\dd Z} \right)^2  \nonumber \\ 
&& +{ 4  \over \gamma \rho} ( \frac{2 \epsilon_{\eta}}{1+2 \epsilon_{\eta}} )
{\dd P\over \dd Z}
\times \left[ {1\over R^3} 
{\dd (R^4 \Omega^2) \over \dd R} \schwz 
- {1 \over R^3} {\dd (R^4 \Omega^2) \over\dd Z} \schwR \right] < 0.
\end{eqnarray}

\noindent For a star with spherically-symmetric isocontours of density, $\rho$,
and pressure, $P$, these criteria take the following form when both
cylindrical ($R$, $Z$) and spherical (${r}$, ${\theta}$) coordinates
are used

\beq
\frac{2 \epsilon_{\eta}}{1+2 \epsilon_{\eta}}N^2 + \kappa^2 > 0,
\eeq
\beq 
(1-\frac{2 \epsilon_{\eta}}{1+2 \epsilon_{\eta}})^2 \left( R 
{\dd \Omega^2 \over\dd Z} \right)^2
-8 ( \frac{2 \epsilon_{\eta}}{1+2 \epsilon_{\eta}}) N^2 \Omega \cot \theta 
\left[ {\dd (\Omega \sin^2 \theta) \over\dd \theta}  \right] < 0.
\eeq

In the limit $\epsilon_{\eta} \rightarrow 0$, these two conditions are
still consistent with the result of Goldreich \& Schubert (1967), even
though these authors focused on the viscous case, rather than the
resistive one. The two necessary conditions for stability are (i) that
the specific angular momentum does not decrease with distance from the
rotation axis and (ii) that the rotation law be constant on
cylinders. The danger in using the limit $\epsilon_{\eta} \rightarrow
0$, as noted by Acheson (1978) for the viscous case, carries over to
the resistive case in the sense that the product $\epsilon_{\eta} N^2
/ \Omega^2$ is not necessarily small compared to unity (see \S4). This
time, the diffusion-free criteria (\S3.1.1) are not exactly recovered
when $\epsilon_{\eta} =1$.

\subsubsection{Requirement $a_4 > 0$ for stability when $\nu \rightarrow 0$}

Like in the perfect-conductor limit, the necessary conditions for
stability are the ``modified Solberg-H\o iland'' criteria derived by
B95 (see \S3.1.3).

\subsubsection{Requirement $a_5 > 0$ for stability when $\nu \rightarrow 0$}

On large enough scales, the requirement $a_5 > 0$ becomes

\beq
-\epsilon_{\eta} \left[ \frac{1}{\gamma \rho}{\cal D}P\> {\cal D}\ln
P\rho^{-\gamma} \right] - \left[
\frac{1}{R^3}{\cal D} (R^4\Omega^2) + 4 \Omega^2 \right] >0.
\eeq

\noindent This condition can be rewritten

\begin{eqnarray}
\left( \frac{k_R}{k_Z} \right)^2 \epsilon_{\eta} N_Z^2+ \frac{k_R}{k_Z} 
\left[ { \epsilon_{\eta} \over \gamma \rho} \left( {\dd P\over \dd Z}\schwR +  
{\dd P\over \dd R}  \schwz \right) - R {\dd \Omega^2 \over\dd Z} \right] 
\nonumber\\
+ \epsilon_{\eta} N_R^2 + {\dd \Omega^2 \over \dd \ln R} >0.  
\end{eqnarray}

Proceeding as before, the two criteria can be written

\beq
\epsilon_{\eta} (N_R^2 + N_Z^2) + \frac{\dd \Omega^2}{\dd \ln R} > 0, 
\eeq

\begin{eqnarray}
(1-\epsilon_{\eta})^2 \left( R {\dd \Omega^2 \over\dd Z} \right)^2
+{ 4 \epsilon_{\eta} \over \gamma \rho} {\dd P\over \dd Z} \left[  
{\dd \Omega^2 \over \dd \ln R} \schwz 
- R {\dd \Omega^2 \over\dd Z} \schwR \right] < 0.
\end{eqnarray}

\noindent For a star with spherically-symmetric isocontours of density, $\rho$,
and pressure, $P$, these criteria take the following form when both
cylindrical ($R$, $Z$) and spherical (${r}$, ${\theta}$) coordinates
are used

\beq
\epsilon_{\eta} N^2 +  \frac{\dd \Omega^2}{\dd \ln R} > 0,
\eeq
\beq 
(1-\epsilon_{\eta})^2 \left( R {\dd \Omega^2 \over\dd Z} \right)^2
-8 \epsilon_{\eta} N^2 \Omega \sin \theta \cos \theta 
\left[ \dd \Omega \over\dd \theta \right] < 0.
\eeq

\noindent In this case, in the limit $\epsilon_{\eta} \rightarrow 0$,
the first of these conditions differs from the result emphasized by
Goldreich \& Schubert (1967): a necessary condition for stability is
that angular velocity (not specific angular momentum) does not
decrease with distance from the rotation axis. The second condition,
requiring the rotation law to be constant on cylinders for marginal
stability, remains the same, however. The diffusion-free result
(\S3.1.3) is recovered for $\epsilon_{\eta}=1$.

\subsubsection{Requirement $det(2) > 0$ for stability when $\nu \rightarrow 0$}

In this limit, on large enough scales, the condition $det(2) > 0$
reduces to the simple expression

\beq
-\left[ \frac{1}{\gamma \rho}{\cal D}P\> {\cal D}\ln
P\rho^{-\gamma}  \right] >0, 
\eeq

\noindent because the various rotational terms cancel out exactly.

This condition can be rewritten

\beq
\left( \frac{k_R}{k_Z} \right)^2 N_Z^2+ \frac{k_R}{k_Z} \left[
{1 \over \gamma \rho} \left( {\dd P\over \dd Z}\schwR +  {\dd P\over \dd R}
 \schwz \right) \right] + N_R^2 > 0.
\eeq

Proceeding as before, the two criteria can be written

\beq
N^2 > 0, \qquad
\left( R \frac{\dd \Omega^2 }{\dd Z} \right)^2 < 0, 
\eeq

\noindent so that marginal stability is possible only for a rotation
law that is constant on cylinders ($\dd \Omega / \dd Z =0$).  Note
that here, the diffusion coefficients $\eta$ and $\chi$ do not
explicitly appear. The necessary condition for the rotation law to be
constant on cylinders is therefore significantly more stringent than
the other $\nu \rightarrow 0$ conditions we have derived thus
far. Because it relies on the exact cancellation of rotational terms
and because it was derived in the inviscid limit, however, this result
may not strictly hold in a more realistic triple-diffusive situation,
when the fluid possesses a finite, even if small, viscosity. We
revisit this issue below when we discuss numerical solutions to our
dispersion relation.

\subsection{Implications}

It is significant that in both the inviscid and perfect-conductor
double-diffusive limits, irrespective of the relative strengths of the
various diffusion processes involved (within the limitations of our
dispersion relation; see Appendix~A), rotation must be constant on
cylinders for stability. In the perfect-conductor limit, this comes
from requiring that $a_5 >0$. In the inviscid limit, this comes from
requiring that the second determinant in the RH analysis be positive
($det(2) > 0$). In both cases, we have also recovered as a necessary
condition for stability the result of Balbus \& Hawley (1994) and B95:
angular velocity must not decrease with distance from the rotation
axis within a given spherical shell for stability in the stellar
context. Consequently, in both double-diffusive limits, a marginally
stable rotation law must at the same time be constant on cylinders and
constant within spherical shells.  This is achieved only by solid body
rotation.\footnote{It is important to note here that our focus is on
"negative" differential rotation, i.e. differential rotation such that
$\dd \Omega / \dd r < 0$ or $\dd \Omega / \dd \theta < 0$ (in terms of
the spherical coordinates $r$ and $\theta$). A "positive" differential
rotation with, for instance, $\dd \Omega / \dd \theta > 0$ and a
rotation constant on cylinders is perfectly stable according to the
criteria we derived.}

We have cautioned that using one or the other double-diffusive limit
can be somewhat misleading because the addition of a third, even if
small, diffusion process should modify the conditions for
stability. Because constant rotation on cylinders is related to
double-diffusive modes of the type described by Goldreich \& Schubert
(1967) or to corresponding magnetized modes in our dispersion
relation, we intuitively expect the double-diffusive results to be
good first approximations as long as focusing on the two largest
diffusion processes is justified. It should be so when there is a well
defined hierarchy of diffusion processes (e.g. $\nu \ll \eta \ll
\xi$). In general, however, it is possible that some finite amount of
stable differential rotation persists in a fully triple-diffusive
situation. We have not been able to establish necessary conditions for
stability in the general, triple-diffusive case and we have addressed
this issue by numerically solving the full dispersion relation.

\section{Numerical Solutions for the Sun's Radiative Zone}

The search for numerical solutions to our dispersion relation
described in this section has three goals: (1) to confirm the results
of our double-diffusive analysis, (2) to explore the potentially
stabilizing role of a third, weaker diffusion process on an otherwise
unstable double-diffusive situation, and (3) to apply our results to
the Sun's radiative zone.

We adopt a standard model for the current Sun (e.g. Bahcall,
Pinsonneault \& Basu 2001; Demarque \& Guenther 1991). We focus on the
part of the radiative zone, from $r \sim 0.3-0.7 R_\odot$, in which
composition gradients are small. We estimate the values of various
microscopic parameters relevant to the stability problem.  Following
Spitzer (1962), the (ion-dominated) dynamic viscosity for a
hydrogen-dominated plasma is

\beq
\mu=\rho \nu \simeq 2.2 \times 10^{-15} \frac{T^{5/2}}{\ln \Lambda}~
{\rm g~cm^{-1}~s^{-1}},
\eeq

\noindent where $\ln \Lambda \sim 4$ is an appropriate value of the Coulomb
logarithm for the Solar interior. The resistivity for a
hydrogen-dominated plasma is

\beq
\eta \simeq 5.2 \times 10^{11} \frac{\ln \Lambda}{T^{3/2}}~
{\rm cm^2~s^{-1}}.
\eeq

Radiative heat diffusion dominates over thermal heat diffusion in the
solar interior, with a radiative conductivity given by (e.g.,
Schwarzschild 1958)

\beq 
\chi_{rad} = {16T^3\sigma\over 3 \kappa\rho}, 
\eeq 

\noindent where $\sigma$ is the Stefan-Boltzmann constant and $\kappa$ is the
radiative opacity. The corresponding radiative diffusivity, which can
be directly compared to the kinematic viscosity and resistivity, is given
by

\beq \xi_{rad}=\frac{\gamma-1}{\gamma} \frac{T}{P} \chi_{rad}.  \eeq

We list in Table~\ref{tab:one} the values of the density, $\rho$,
temperature, $T$, Rosseland-mean opacity, $\kappa$ (obtained from a
standard opacity table) and all three diffusivities over the region of
interest in the solar radiative zone. In addition, the values of the
Prandtl and Acheson numbers, $\epsilon_\nu$ and $\epsilon_\eta$, which
are directly relevant to the stability analysis, are listed. Clearly,
resistive diffusion dominates over viscous diffusion in the bulk of
the solar radiative zone, by a factor $\sim 20-30$. Note that the
radiative kinematic viscosity

\beq
\nu_r=\frac{16}{15} \frac{\sigma T^4} {\kappa \rho^2 c^2},
\eeq

\noindent where $c$ is the speed of light, makes only a small contribution to
the total kinematic viscosity in that region (e.g. Goldreich \&
Schubert 1967). In our specific numerical applications, we focus on
the region below the convection zone at $r \ltsim 0.7$, for which
helioseismological measurements of a near solid-body rotation are most
reliable, with an angular velocity $\Omega =2.7 \times
10^{-6}$~rad~s$^{-1}$ (see, e.g., Charbonneau, Dikpati \& Gilman
1999). We adopt the value $N^2 = 1.3 \times 10^{-6}$~Hz$^2$ for the
square of the \BV frequency (Demarque \& Guenther 1991; their
Fig. 23), so that the ratio $\Omega^2 / N^2 = 5.6 \times 10^{-6}$ in
our models.

It is important to note at this point that, in the Sun's radiative
zone, the Prandtl number, $\epsilon_\nu$, is comparable to $\Omega^2/
N^2$ while the Acheson number, $\epsilon_\eta$, is systematically $\gg
\Omega^2/ N^2$.  Indeed, Acheson (1978) has pointed out that, since
the double-diffusive stability criteria of Goldreich \& Schubert
(1967) are typically of the form $\epsilon_\nu N^2 + \kappa^2 >0$ (see
\S3.1.2), substantial differential rotation, for instance a
Keplerian-like profile with $\kappa^2=- \Omega^2$, can remain stable
depending on the exact values of $\epsilon_\nu$, $\Omega^2$ and $N^2$.
This objection could be a practical limitation of the results of
Goldreich \& Schubert (1967; see also Fricke 1968), who focused on the
idealized limit $\epsilon_\nu \to 0$. Acheson's point is made even
stronger by observing that the relevant double-diffusive limit for the
Sun is the inviscid one (since $\chi \gg \eta > \nu$) and that
$\epsilon_\eta \gg \Omega^2/N^2$ (see Table~\ref{tab:one}), which
appears to be strongly stabilizing by Acheson's arguments.  Our
extensive stability analysis in the previous section shows, however,
that in both the inviscid and perfect-conductor double-diffusive
limits, any level of negative differential rotation is destabilized by
a combination of diffusion-free and double-diffusive modes,
irrespective of the relative strength of viscous, resistive and heat
diffusion. This result effectively invalidates Acheson's objection but
it remains to be seen what the effects on stability of adding a third,
weaker diffusion process may be.

We have numerically solved the complete dispersion relation
(Eq.~[\ref{poly}]) with parameters appropriate for the Sun's upper
radiative zone, using the Laguerre algorithm described by Press et al.
(1992). For definiteness, we have considered only conditions
appropriate to the specific radius $r \simeq 0.7 R_\odot$: $\nu_\odot
= 23.6$ cm$^2$~s$^{-1}$, $\eta_\odot=596$ cm$^2$~s$^{-1}$ and
$\xi_{rad \odot} = 1.2 \times 10^7$ cm$^2$~s$^{-1}$. In general, we
considered a range of values for the polar angle, $\theta$, in the
interval $[0,\pi/2]$ (pole to equator). We have found it useful to
rewrite both the rotational and stratification terms appearing in the
coefficients $a_2$--$a_5$ of the dispersion relation in terms of
spherical coordinates $(r, \theta, \phi)$. For a spherical star, the
stratification term becomes only function of $N^2$, $\theta$ and the
radial and angular wavevectors, $k_r$ and $k_\theta$.  The rotational
term, on the other hand, explicitly depends on the amount of
differential rotation within and between spherical shells, that we
express as $\dd \ln \Omega / \dd \theta$ and $\dd \ln \Omega / \dd \ln
r$, respectively. As we have mentioned earlier, we are mostly
interested in negative differential rotation, i.e. cases where $\dd
\ln \Omega / \dd \theta < 0$ and/or $\dd \ln \Omega / \dd \ln r < 0$.
We will consider these two cases separately because we expect
differential rotation within spherical shells to be destabilized by
diffusion-free modes and differential rotation between shells to be
destabilized by multi-diffusive modes.

When searching for unstable modes, we vary the wavevectors $k_r$ and
$k_\theta$ independently in the range 

\begin{eqnarray} 
\pm \frac{2 \pi}{10^{-2} R_\odot} \to \pm \frac{2 \pi}{10^{-14}
R_\odot}. \nonumber
\end{eqnarray} 

\noindent The first (large-scale) limit guarantees that we are looking
at scales significantly smaller than the pressure scale height
($H_\odot \sim 0.1 R_\odot$), while the second (small scale) limit
guarantees that we are looking at scales significantly in excess of
the mean free path for the conditions of interest.  We have also
performed focused searches on small regions of the wavevector space
corresponding to nearly cylindrical--radial ($|k_R/k_Z| \to 0$)
displacements or nearly vertical ($|k_R/k_Z| \to \infty$)
displacements. Independently of the values of the wavenumbers, we have
varied the Alfv\`en speeds $v_{Ar}$ and $v_{A \theta}$ in the range
$10^{-2}$ -- $10^{-26} R_\odot \Omega$ and we have also explored cases
with $v_{Ar}=0$ and $v_{A \theta}=0$. This large range of values for
the Alfv\`en speeds, independently of the values for $k_r$ and
$k_\theta$, effectively provides a search both in magnetic field
strength and geometry. With this extensive search, we have explored
regimes in which $|\kva|$ is successively $\gg$, $\ll$ and comparable
to each of the three dissipation terms ($\nu k^2$, $\eta k^2$ and
$\xi_{rad} k^2$).

In the inviscid limit ($\nu = 0$), we were able to find unstable modes
down to values of differential rotation between shells as low as $\dd
\ln \Omega / \dd \ln r \ltsim -0.1$ to $-0.01$, depending on the value
of the polar angle, $\theta$. In the perfect-conductor limit ($\eta
=0$), we were able to find unstable modes down to values as low as
$\dd \ln \Omega / \dd \ln r \leq -0.01$ for essentially all polar
angles. All the unstable modes were of the direct type
($Im(\sigma)=0$), as opposed to the overstable type ($|Im(\sigma)| >
Re(\sigma)>0$), as expected in the situation of interest, with a
strongly stabilizing thermal stratification and a fast rate of heat
diffusion. That we found unstable modes down to such very low values
of negative differential rotation between shells in both
double-diffusive limits is consistent with the conclusions of our
stability analysis in \S3.

We then turned to the fully triple-diffusive situation and first
investigated differential rotation within spherical shells (i.e. $\dd
\ln \Omega / \dd \theta < 0$ only). We expect this type of
differential rotation to be destabilized by diffusion-free modes,
which should be easily identified even in the triple-diffusive
case. Indeed, we have found that unstable modes were easily identified
down to levels $\dd \ln \Omega / \dd \theta \ltsim -0.1$ to $-0.01$,
depending on the value of the polar angle, $\theta$.

We then searched for unstable modes in the presence of differential
rotation between shells (i.e. $\dd \ln \Omega / \dd \ln r <0$
only). We have been able to identify unstable modes down to levels of
differential rotation corresponding to $\dd \ln \Omega / \dd \ln r
\simeq -1.2$ to $-1.5$, depending on the value of the polar angle,
$\theta$. It became clear during the extensive search required to
identify these triple-diffusive modes that they are located in a much
smaller region of the parameter space than the unstable modes we
previously identified in the corresponding double-diffusive
limit. This is as expected if the addition of a third, weaker
diffusion process stabilizes an otherwise unstable double-diffusive
situation. We have confirmed this stabilizing effect explicitly by
observing that an initially weakly unstable triple-diffusive situation
slowly makes a transition to stability as the value of the third,
weakest diffusion coefficient is increased.

In that respect, it is worth noting that the triple-diffusive
situation in the Sun is such that the weakest diffusion process,
viscosity, is ``only'' a factor $20$--$30$ times smaller than the
second weakest one, resistivity, and this may be a source of
stabilization with respect to small levels of differential rotation
between shells. During our numerical exploration of the
triple-diffusive situation in the Sun's upper radiative zone, we have
also noticed a rather strong sensitivity of the stability results to
small variations (say, $\times 2$) in the values of parameters such as
the viscosity, $\nu_\odot$, or the \BV frequency, $N$. A complete
stability analysis for the current Sun is beyond the scope of the
present study.  It is encouraging, nonetheless, that we have been able
to identify unstable modes in the presence of moderate levels of
differential rotation for the current upper radiative zone conditions,
as it suggests that these modes may have played an important role in
establishing the current rotation profile in the Sun.

\section{The Emergence of Solid Body Rotation}

It has long been realized that the timescale for microscopic viscosity
to reduce significant levels of differential rotation in the Sun is
excessively long ($R_\odot^2 / \nu_\odot \sim 0.5$-$1 \times
10^{13}$~years; see, e.g., Goldreich \& Schubert 1967). On the other
hand, differential rotation (with $\dd \ln \Omega / \dd \ln r < 0$) is
expected to have been present in the early Sun, because of the likely
fast initial rotation and of the magnetic spin-down torque externally
exerted via the solar wind (see, e.g., Sofia et al. 1991 for a
review).  Consequently, the solid body rotation inferred for the Sun's
upper radiative zone from seismology (e.g., Kosovichev et al. 1997;
Schou et al. 1998; Charbonneau et al. 1999) requires a mechanism
capable of reducing differential rotation much more efficiently.

Several mechanisms have been proposed and discussed at length in the
literature (see, e.g., Schatzman 1991 for a review).  They include the
Goldreich-Schubert-Fricke (GSF) instability, the ``secular
hydrodynamical shear instability'' (Zahn 1974) and angular momentum
transport by internal gravity waves (Press 1981; Kumar \& Quataert
1997; Talon \& Zahn 1998; Talon, Kumar \& Zahn 2002). None of these
mechanisms may be able to provide a satisfactory solution, however.
The GSF instability alone is not expected to bring a system to a state
of solid body rotation (but only of rotation constant on cylinders;
e.g., Goldreich \& Schubert 1967). The existence of a secular shear
instability has not been rigorously proven, but only inferred from
heuristic arguments (Zahn 1974; Schatzman 1991), and it would operate
in the current Sun's radiative zone only in the presence of strong
shear (Zahn 1993). Finally, while internal gravity waves may transport
angular momentum efficiently, they are not generally expected to be
efficient at mixing elements. The large amount of Li depletion at the
surface of the Sun and other stars (see, e.g., Chaboyer, Demarque \&
Pinsonneault 1995a,b) is best interpreted as resulting from turbulent
mixing in stellar radiative zones, thus favoring the action of
instabilities rather than waves to explain both mixing and rotational
evolution.

According to our analysis, it is the combination of weak magnetic
fields and multi-diffusive modes that may render even a small level of
(negative) differential rotation unstable. Independently of this issue
of stability, there are two important aspects of the problem that a
linear analysis cannot address. First, it is {\it a priori} unclear
whether the turbulence resulting from the non-linear development of
the various modes described by our dispersion relation will drive the
system towards a state of marginal stability (i.e. near solid body
rotation). Intuitively, because the main force balance in a star does
not involve rotation (contrary to an accretion disk), we would expect
the turbulence to be relatively ``free'' to bring a star close to a
state of marginal stability, but this remains to be proven.  Second,
it is also unclear with what efficiency (i.e. on what timescale) that
turbulence would be able to transport angular momentum and affect an
unstable rotation profile. Addressing these two non-linear aspects of
the problem reliably would require fully turbulent numerical
simulations. In that respect, it is interesting to note that a
preliminary two-dimensional investigation by Korycansky (1991) of
hydrodynamical double-diffusive modes in a specific "equatorial"
geometry does indicate that angular momentum transport drives the
system towards marginal stability. An additional motivation for
carrying out such detailed numerical simulations of turbulence driven
by multi-diffusive, magnetized and unmagnetized modes would be to
estimate the efficiency of turbulent transport of elements.

One important physical element that has been neglected in our analysis
is the stabilizing effect of composition gradients. Goldreich \&
Schubert (1967) have described, in the single-diffusive hydrodynamical
limit, how even moderate gradients of chemical composition can
stabilize significant levels of radial differential rotation. As these
authors have noted, this effect would be important at radii $r \ltsim
0.3 R_\odot$ in the Sun's core, a region that we have excluded from
our analysis. While a complete derivation of our dispersion relation
including the effects of composition gradients is beyond the scope of
the present study, by analogy with the results of Goldreich \&
Schubert (1967), one may expect significantly stronger levels of
radial differential rotation to be maintained in the solar core,
relative to the rest of the radiative zone. On the other hand, it is
possible that the early differential rotation in the region currently
encompassing the solar core was reduced before significant hydrogen
burning took place, if multi-diffusive modes were rather efficient
early on at redistributing angular momentum. This illustrates how the
problem of differential rotation is intimately linked to that of
stellar evolution.

\section{Conclusion}

We have studied the local axisymmetric triple-diffusive stability of
stratified, weakly-magnetized, differentially-rotating fluids. We have
established that, in an inviscid or a perfectly-conducting fluid,
differential rotation is destabilized by a combination of
diffusion-free and double-diffusive modes, unless rotation is constant
on cylinders and angular velocity does not decrease away from the
rotation axis. We have stressed the important role of weak magnetic
fields in establishing these results. We have found that, in a more
realistic triple-diffusive situation, the weakest diffusion process
can sometimes play a stabilizing role. While our analysis is rather
general, we have discussed a specific numerical application to the
Sun's upper radiative zone, which is seismologically known to be
rotating near solid body rotation. We have found that moderate to
strong levels of differential rotation, if present, would indeed be
destabilized, thus suggesting that magnetized and multi-diffusive
modes may have played an important role in establishing the current
solar internal rotation.

\section*{Acknowledgments}

KM gratefully acknowledges support from the Celerity Foundation.
Support for this research was provided by NASA Grants NAG5--13288 and
NAG5-9266.  

\section*{Appendix A: Limitations}

\subsection*{A1: Validity of the dispersion relation}

{We neglected the resistive and viscous dissipation terms in the
entropy equation (Eq.~[4]; see, e.g., Tassoul 1978 or Balbus \& Hawley
1998 for complete formulations) and its subsequent linearized form
(Eq.~[\ref{lincon}])}.  We explore the range of validity of this
approximation here.

Let us first compare the magnitudes of the three terms we kept in
equation~(\ref{lincon}) for entropy perturbations. The ratio of the
first term on the LHS to the perturbed heat diffusion term is of order

\beq
R_1=\frac{\omega \delta T/T}{\chi k^2 \delta T / P} \sim \frac{\omega}{\chi k^2 T/P} \sim 
\frac{\omega}{(\xi/\nu) \lambda c_s k^2} \sim \frac{1}{(\xi/\nu)(kH)(k \lambda)},
\eeq

\noindent where we have used the definition of the heat diffusivity,
$\xi$, and we have equated the kinematic viscosity coefficient, $\nu$,
to the product of the mean free path, $\lambda$, and sound speed,
$c_s$. Note that the perturbation frequency, $\omega$, is typically
(though not exclusively) $\sim N$ (the \BV frequency) in a
pressure-supported system and $\sim \Omega$ (the rotation frequency)
in a rotation-supported system. In both cases, $\omega \sim c_s/H$,
where $H$ is the pressure scale-height of the system. Irrespective of
the ratio of diffusivities ($\xi/\nu$), our local analysis making use
of the MHD equations is valid only for scales much smaller than the
system's scale-height ($kH \gg 1$) and much larger than the mean free
path ($k \lambda \ll 1$). In general, the ratio $R_1$ can therefore be
$\gg 1$ or $\ll 1$ and one must keep the heat diffusion term in the
perturbed entropy equation to leading order.

The ratio of the same first term on the LHS of equation~(\ref{lincon})
to the (neglected) perturbed viscous heating term is of order

\beq
R_2= \frac{\omega \delta T/T}{\delta[\mu \left| {d\Omega/d\ln r} \right|^2 ]/P} \sim
\frac{\omega \delta v / c_s}{\nu \Omega k \delta v /c_s^2} \sim \frac{\omega}{k \lambda \Omega},
\eeq

\noindent where we have used the relations $\delta v / c_s \sim \delta \rho /
\rho \sim \delta T / T$ and $| {d\Omega/d\ln r } | \sim
\Omega$. Since $\omega / \Omega$ is typically $\sim 1$
(rotation-supported system) or $\gg 1$ (pressure-supported system),
and $k \lambda \ll 1$, $R_2$ is $\gg 1$ and neglecting the perturbed
viscous heating term is justified to leading order.
 
Similarly, the ratio of the first term on the LHS of equation~(\ref{lincon}) 
to the (neglected) perturbed resistive heating term is of order
 
 \beq
 R_3=\frac{\omega \delta T / T}{\delta[(\eta/4 \pi) \left| {\bb \nabla\times B} \right|^2]/P} \sim
\frac{\omega \delta T / T}{(\eta/4 \pi)(v_A/c_s)^2 k \delta c_s /(c_s H)} \sim \frac{1}{(\eta / 8
\pi \nu) (v_A/c_s)^2 (k \lambda)},
 \eeq
 
\noindent where we have made the additional assumption that the ratio $v_A/c_s$
is locally constant in the basic state configuration. This amounts to
requiring that the basic state magnetic field does not possess strong
gradients on scales smaller than the scale height, $H$, and is a
reasonable approximation unless one is interested in rather singular
basic state magnetic field configurations. Irrespective of the ratio
of diffusivities, $\eta/\nu$, our weak field assumption implies that
$v_A/c_s \ll 1$, so that $R_3$ is also $\gg 1$ in general and
neglecting the perturbed resistive heating term is justified to
leading order.

Physically, this hierarchy of heating terms in the perturbed entropy
equation can be understood by noting that heat diffusion is naturally
the most efficient form of heat transport, unless the viscosity and/or
resistivity coefficients are very much larger than the heat
diffusivity coefficient (i.e. $\nu$ or $\eta \gg \xi$). That our
assumptions for the leading-order perturbed equations remain valid
even if $\nu$ or $\eta$ is larger than $\xi$, but not so much as to
become more efficient at transporting heat than heat diffusion itself,
indicates that our dispersion relation is able to describe overstable
modes in a strongly stratified medium where such a hierarchy of
diffusivities occurs.

\subsection*{A2: How weak can weak magnetic fields be?}

In the coefficients $a_4$ and $a_5$ of our dispersion relation,
pre-factors multiplying the rotational and stratification terms
involve sums of dissipation and magnetic tension terms. Our
double-diffusive stability analysis in \S3 assumes that the magnetic
field strength, while weak ($v_A \ll c_s $ and $v_A \ll (HR)^{1/2}
\Omega$), is strong enough for the magnetic tension term to dominate
over the dissipation terms in all these pre-factors. We determine the
range of validity of this assumption here, using numerical values
appropriate for the Sun.

Let $\xi$ denote the largest diffusivity coefficient (in units of
cm$^2$~s$^{-1}$) in the medium of interest. For the assumption made in
our double-diffusive stability analysis to be incorrect, we need $k^2
v_A^2 \ll \xi^2 k^4$ for all possible values of $k$ relevant to the
local analysis. Since the minimum acceptable value of $k$ is $\sim 2
\pi/ H_\odot$ (where $H_\odot$ is the scale height), our assumption
breaks down for all relevant scales if $v_A^2 \ll 4 \pi^2
\xi^2/H_\odot^2$. Using typical values $\rho \sim 1$~g~cm$^{-3}$, $\xi
\sim 10^7$~cm$^2$~s$^{-1}$ and $H_\odot / R_\odot \sim 0.1$ for the
Sun's upper radiative zone, this translates into a limit on the field
strength $B \ll 8 \times 10^{-5}$~G, a very small value indeed.  Since
the above-mentioned pre-factors in $a_4$ and $a_5$ actually involve
products of several diffusivity coefficients (not just the largest
one), the limit on the field strength will be smaller than we
estimated by several extra orders of magnitude for the conditions in
the Sun's upper radiative zone.  On the other hand, our assumption of
weak magnetic field leads to $v_A^2=B^2/4\pi \rho \ll (H_\odot
R_\odot)^{1/2} \Omega \ll c_s^2$ or $B \ll 10^3 $~G. This still leaves
a comfortable range of field strengths for which the assumptions made
in our double-diffusive stability analysis are valid for the solar
interior.

\clearpage

\begin{table}[htdp]
\caption{Diffusive Conditions in the Solar Radiative Zone}
{\footnotesize
\begin{center}
\begin{tabular}{cccccccccc}
\hline
\\
Radius & $\rho$&$T$&$\kappa$&$\nu$&$\nu_r$&$\eta$&$\xi_{rad}$&$\epsilon_{\nu}$&$\epsilon_{\eta}$\\
$(R_\odot)$&(g cm$^{-3}$)&($10^6$ K)&(cm$^2$ g$^{-1}$)&(cm$^2$ s$^{-1}$)&(cm$^2$ s$^{-1}$)&(cm$^2$
s$^{-1}$)&(cm$^2$ s$^{-1}$)&&\\
\\
\hline
\\
$r \simeq 0.7$& $0.2$& $2.3$ & $18$&$21$&$2.6$&$596$&$1.2 \times 10^7$&$2 \times 10^{-6}$&$5 \times
10^{-5}$\\
$r \simeq 0.3$& $8.5$ & $6.2$ & $3$&$6$&$0.5$&$135$&$8.1 \times 10^5$ & $8 \times 10^{-6}$ & $1.7 \times
10^{-4}$\\
\\
\hline
\end{tabular}
\end{center}}
\label{tab:one}
\end{table}%


\begin{references}

\reference{} Acheson, D.~J. 1978, Phil. Trans. Roy. Soc. London, 289A, 459

\reference{} Armitage, P.~J. 1998, ApJ, 501, L189

\reference{} Bahcall, J.~N., Pinsonneault, M.~H. \& Basu, S. 2001, ApJ, 555, 990

\reference{} Balbus, S.~A. \& Hawley, J.~F. 1991, ApJ, 376, 214 

\reference{} Balbus, S.~A. \& Hawley, J.~F. 1994, MNRAS, 266, 769

\reference{} Balbus, S.~A., \& Hawley, J.~F. 1998, Rev.~Mod.~Phys., 70, 1

\reference{} Balbus, S.~A. 1995, ApJ, 453, 380 (B95)

\reference{} Balbus, S.~A. 2001, ApJ, 562, 909 

\reference{} Balbus, S.~A. 2003, ARA\&A, in press (preprint)

\reference{} Blaes, O. 2003, in "Accretion Disks, Jets, and High
Energy Phenomena in Astrophysics", Proceedings of Session LXXVIII of
Les Houches Summer School, August 2002, eds. F. Menard, G. Pelletier,
G. Henri, V. Beskin, and J. Dalibard (EDP Science: Paris and Springer:
Berlin), astro-ph/0211368

\reference{} Chaboyer, B., Demarque, P. \& Pinsonneault, M. H. 1995a,
ApJ, 441, 865

\reference{} Chaboyer, B., Demarque, P. \& Pinsonneault, M. H. 1995b,
ApJ, 441, 876

\reference{} Charbonneau, P., Dikpati, M. \& Gilman, P.~A. 1999, ApJ, 526, 523

\reference{} Demarque, P. \& Guenther, D. B. 1991, in Solar Interior
and Atmosphere, eds. A.N. Cox, W.C. Livingston and M.S. Matthews
(Tucson: University of Arizona Press), p. 1186

\reference{} Fricke, K. 1968, Z. Ap., 68, 137

\reference{} Goldreich, P. \& Schubert, G. 1967, ApJ, 150, 571

\reference{} Hawley, J.~F. 2001, ApJ, 554, 534

\reference{} Hawley, J.~F., Gammie, C.~F. \& Balbus, S.~A. 1995, ApJ, 440, 742  
\reference{} Korycansky, D.~G. 1991, ApJ, 381, 515

\reference{} Kosovichev, A.~G. et al. 1997, Sol Phys., 170, 43

\reference{} Kumar, P. \& Quataert, E.~J. 1997, ApJ, 479, L51 


\reference{} Parker, E.~N. 1979, Cosmical Magnetic Fields (New York: Oxford University Press)

\reference{} Press, W.~H., Teukolsky, S.~A., Vetterling, W.T. \& Flannery, B.~P. 1992, Numerical
Recipes in Fortran. The art of scientific computing (Cambridge: Cambridge University Press)

\reference{} Schou, J. et al. 1998, ApJ, 505, 390

\reference{}
Schwarzschild, M. 1958, Structure and Evolution of the Stars (New York:
Dover)

\reference{} Sofia, S., Kawaler, S., Larson, R. \& Pinsonneault,
M. 1991, in Solar Interior and Atmosphere, eds. A.N. Cox,
W.C. Livingston and M.S. Matthews (Tucson: University of Arizona
Press), p. 140

\reference{} Schatzman, E. 1991, in Solar Interior and Atmosphere,
eds. A.N. Cox, W.C. Livingston and M.S. Matthews (Tucson: University
of Arizona Press), p. 192

\reference{} Spiegel, E.~A. \& Zahn, J.~P. 1992, A\&A, 265, 115

\reference{}
Spitzer, L. 1962, Physics of Fully Ionized Gases (New York: Wiley
Interscience)

\reference{} Talon, S. \& Zahn, J.-P. 1998, A\&A, 329, 315

\reference{} Talon, S., Kumar, P. \& Zahn, J.-P. 2002, ApJ, 574, L175

\reference{}
Tassoul, J.-L. 1978, Theory of Rotating Stars (Princeton: Princeton
University Press)

\reference{} Urpin, V. \& Brandenburg, A. 1998, MNRAS, 294, 399

\reference{} Zahn, J.-P. 1974, in Stellar Instability and Evolution,
Proc. IAU Symp. 129, eds. P. Ledoux, A. Noels and A.W. Rodgers (Dordrecht: D. Reidel), p. 185

\reference{} Zahn, J.-P. 1993, in Les Houches School (Session XLVII):
Astrophysical Fluid Dynamics, eds. J.-P. Zahn \& J. Zinn-Justin
(Elsevier), p. 561

\end{references}
\end{document}